\documentclass[fleqn,usenatbib]{mnras}

\usepackage{newtxtext,newtxmath}

\usepackage[T1]{fontenc}

\DeclareRobustCommand{\VAN}[3]{#2}
\let\VANthebibliography\thebibliography
\def\thebibliography{\DeclareRobustCommand{\VAN}[3]{##3}\VANthebibliography}

\usepackage{graphicx}	
\usepackage{amsmath}	
\usepackage{cases}

\newcommand{\B}[1]{\langle B_{\rm #1} \rangle}
\newcommand{\BE}[1]{\langle B_{\rm #1}^2 \rangle}

\title[Slowly rotating fully convective M dwarfs]{The magnetic and spin-down properties of slowly rotating fully convective M dwarfs}

\author[V. See et al.]{Victor See$^{1,2}$\thanks{E-mail: w.c.v.see@bham.ac.uk},
Louis Amard$^{3,4}$,
Stefano Bellotti$^{5,6}$,
Sudeshna Boro Saikia$^{7}$,
Emma L. Brown$^{8,9}$, \newauthor
Jean-Francois Donati$^{6}$,
Rim Fares$^{10}$,
Adam J. Finley$^{3}$,
Colin P. Folsom$^{11}$,
\'Elodie M. H\'ebrard, \newauthor
Moira M. Jardine$^{12}$,
Sandra V. Jeffers$^{13}$, 
Baptiste Klein$^{14}$,
Lisa T. Lehmann,
Stephen C. Marsden$^{8}$, \newauthor
Sean P. Matt$^{15}$,
Matthew W. Mengel$^{8}$, 
Julien Morin$^{16}$, 
Pascal Petit$^{6}$,
Katelyn Smith$^{17,5}$, \newauthor
Aline A. Vidotto$^{5}$,
Ian A. Waite$^{8}$.
\\
$^{1}$School of Physics \& Astronomy, University of Birmingham, Edgbaston, Birmingham B15 2TT, UK\\
$^{2}$European Space Agency (ESA), European Space Research and Technology Centre (ESTEC), Keplerlaan 1, 2201 AZ Noordwijk, The Netherlands\\
$^{3}$Universit\'e Paris-Saclay, Universit\'e Paris Cit\'e, CEA, CNRS, AIM, 91191, Gif-sur-Yvette, France\\
$^{4}$Department of Astronomy, University of Geneva, Chemin Pegasi 51, 1290 Versoix, Switzerland\\
$^{5}$Leiden Observatory, Leiden University, PO Box 9513, 2300 RA Leiden, The Netherlands\\
$^{6}$IRAP, Universit\'e de Toulouse, CNRS/UMR 5277, UPS-OMP, 14 Avenue E. Belin, Toulouse F-31400 France\\
$^{7}$University of Vienna, Department of Astrophysics, Turkenschanzstrasse 17, 1180 Vienna, Austria\\
$^{8}$Centre for Astrophysics, University of Southern Queensland, Toowoomba, QLD 4350, Australia\\
$^{9}$School of Access Education, Central Queensland University, Cairns, QLD 4870, Australia\\
$^{10}$Physics Department, United Arab Emirates University, P.O. Box 15551, Al-Ain, United Arab Emirates\\
$^{11}$Tartu Observatory, University of Tartu, Observatooriumi 1, T\~oravere, 61602, Estonia\\
$^{12}$SUPA, School of Physics and Astronomy, University of St Andrews, North Haugh, St Andrews KY16 9SS, UK\\
$^{13}$Th\"uringer Landessternwarte Tautenburg, Sternwarte 5, D-07778 Tautenburg, Germany\\
$^{14}$Department of Physics, University of Oxford, OX1 3RH, Oxford, UK\\
$^{15}$Homer L. Dodge Department of Physics and Astronomy, University of Oklahoma, Norman, OK 73019, USA\\
$^{16}$Laboratoire Univers et Particules de Montpellier, Universit\'e de Montpellier, CNRS, F-34095, Montpellier, France\\
$^{17}$School of Mathematics and Physics, University of Queensland, St Lucia, QLD 4072, Australia
}

\date{Accepted XXX. Received YYY; in original form ZZZ}

\pubyear{2023}

\begin{document}
\label{firstpage}
\pagerange{\pageref{firstpage}--\pageref{lastpage}}
\maketitle

\begin{abstract}
The evolution of the magnetism, winds and rotation of low-mass stars are all linked. One of the most common ways to probe the magnetic properties of low-mass stars is with the Zeeman-Doppler imaging (ZDI) technique. The magnetic properties of partially convective stars has been relatively well explored with the ZDI technique, but the same is not true of fully convective stars. In this work, we analyse a sample of stars that have been mapped with ZDI. Notably, this sample contains a number of slowly rotating fully convective M dwarfs whose magnetic fields were recently reconstructed with ZDI. We find that the dipolar, quadrupolar and octupolar field strengths of the slowly rotating fully convective stars do not follow the same Rossby number scaling in the unsaturated regime as partially convective stars. Based on these field strengths, we demonstrate that previous estimates of spin-down torques for slowly rotating fully convective stars could have been underestimated by an order of magnitude or more. Additionally, we also find that fully convective and partially convective stars fall into distinct sequences when comparing their poloidal and toroidal magnetic energies.
\end{abstract}




\section{Introduction}
Dynamo action in the interiors of low-mass stars ($M_\star \lesssim 1.3M_\odot$) is responsible for the generation of stellar magnetic fields and their associated magnetic activity. Although the precise mechanisms that drive dynamo action are still not well understood, it is thought that the interaction between a star's rotation and convective motions plays a crucial role \citep{Brun2017,Kapyla2023}. To first order, the interplay between these motions can be captured by a dimensionless parameter known as the stellar Rossby number, defined as the ratio of the rotation period over the convective turnover time, ${\rm Ro}=P_{\rm rot}/\tau$. Many studies have shown that stellar magnetic field strengths and magnetic activity scale with the stellar Rossby number \citep{Noyes1984,Saar1999,Pizzolato2003,Mamajek2008,Reiners2009,Vidotto2014,See2015,Stelzer2016,Newton2017,Wright2018,Folsom2018,See2019ZBvsZDI,See2019Geom,Kochukhov2020,Reiners2022,Boudreaux2022,Cao2022,Cristofari2023,Gossage2024}. In general, stellar magnetism and activity can be divided into two regimes known as the unsaturated and saturated regimes. In the unsaturated regime, magnetic activity increases with decreasing Rossby number until a critical Rossby number, $\rm Ro_{crit}$, is reached. Stars with Rossby numbers smaller than this critical Rossby number are in the saturated regime and their activity has a much flatter dependence on the Rossby number or is even independent of the Rossby number. 

One direct consequence of stars having magnetic fields is the existence of magnetised stellar winds. These are outflows of ionised gas that carry angular momentum away from stars \citep[e.g.][]{Mestel1968}. As a result, stellar rotation slows over time which can be seen in rotation period measurements of stars in open clusters of different ages \citep[e.g.][]{Meibom2009, Meibom2011, Barnes2016,Rebull2016,Douglas2016,Douglas2017,Douglas2019,Curtis2020,Fritzewski2021,Dungee2022}. Although it is not possible to directly measure the rate at which stars lose angular momentum through their winds \citep[with the exception of the Sun;][]{Finley2019,Finley2020,Finley2021}, it is expected to be a function of the Rossby number \citep[e.g.][]{vanSaders2013,Matt2015,Johnstone2021,Breimann2021}. As such, there is a feedback loop where the rate of angular momentum loss is dependent on the star's rotation (via the Rossby number) and the star's rotation is being constantly slowed by the angular momentum loss.

Although stellar rotation and magnetism have been extensively studied, there still remain open questions. One such question is whether the same dynamo processes are responsible for magnetic field generation within fully convective (FC) and partially convective (PC) stars. Given that FC stars do not posses a tachocline, which is thought to play a role in generating the magnetic fields of PC stars, one might expect different types of dynamos to operate in FC and PC stars resulting in FC and PC stars having different magnetic and spin-down properties. Indeed, a number of findings suggest this could be the case. The first is that the magnetic field properties of rapidly rotating FC stars appear to be bimodal; either strong and dipolar or weak and multipolar \citep{Morin2010}. This bimodality in magnetic properties for stars that are otherwise similar has not yet been observed in other types of stars and hints that the dynamos of FC stars may be different in some sense. Another piece of evidence that the dynamos of FC and PC stars operate differently is the recent finding that FC stars seem to lose angular momentum through their winds more rapidly than PC stars, even when other properties, such as mass, rotation period, etc., are similar \citep{Lu2024}. This should not occur if the magnetic properties of FC and PC stars scale in the same way. However, in contrast to the findings of these investigations, other studies have shown that the X-ray emission from FC and PC stars follow the same scaling relation as a function of Rossby number \citep{Wright2016,Wright2018,Gossage2024} suggesting that the dynamos of FC and PC stars may be operating similarly. Currently, it remains unclear why some studies suggest that different dynamo processes are in operation within FC and PC stars while other studies suggest that their dynamos are similar.

In this work, we will further explore the question of whether FC and PC stars have different types of dynamos. We do this by analysing a sample of stars that have had their magnetic fields reconstructed with Zeeman-Doppler imaging (ZDI). ZDI is a tomographic imaging technique that can reconstruct the large-scale magnetic field strength and geometry at the photospheres of stars \citep{Semel1989,Brown1991,Donati1997,Donati2006}. Previous studies have already used samples of ZDI maps to investigate how the magnetic and spin-down properties of stars vary over a wide range of spectral types and rotation rates \citep{Donati2008,Morin2008,Morin2010,Vidotto2014,See2015,See2016,See2019ZBvsZDI,See2019Geom,Finley2019Stellar,Folsom2016,Folsom2018}. However, the samples in these studies did not contain any slowly rotating FC stars as no ZDI maps existed for these types of stars at the time. As such, the magnetic and spin-down properties of slowly rotating FC stars are less well understood. This situation has recently changed with newly published ZDI maps of FC stars \citep{Klein2021,Lehmann2024}. These investigations show that the large-scale magnetic fields of slowly rotating FC stars can be stronger than PC stars with comparable Rossby numbers which, again, is suggestive that different dynamo processes are operating within FC and PC stars. In this work, we will use these new ZDI maps to further investigate the magnetic and spin-down properties of slowly rotating FC stars.

The rest of this paper is outlined as follows. We cover the details of our sample in section \ref{sec:Sample}. In section \ref{sec:MagProp}, we look at the magnetic properties of the stars in our sample paying special attention to the slowly rotating FC stars. We analyse the spin-down properties of our stars in section \ref{sec:Spindown}. In section \ref{sec:Discussion}, we discuss some caveats and implications of our work and we summarise our conclusions in section \ref{sec:Conclusion}.

\begin{figure}
	\includegraphics[trim=0mm 10mm 0mm 0mm,width=\columnwidth]{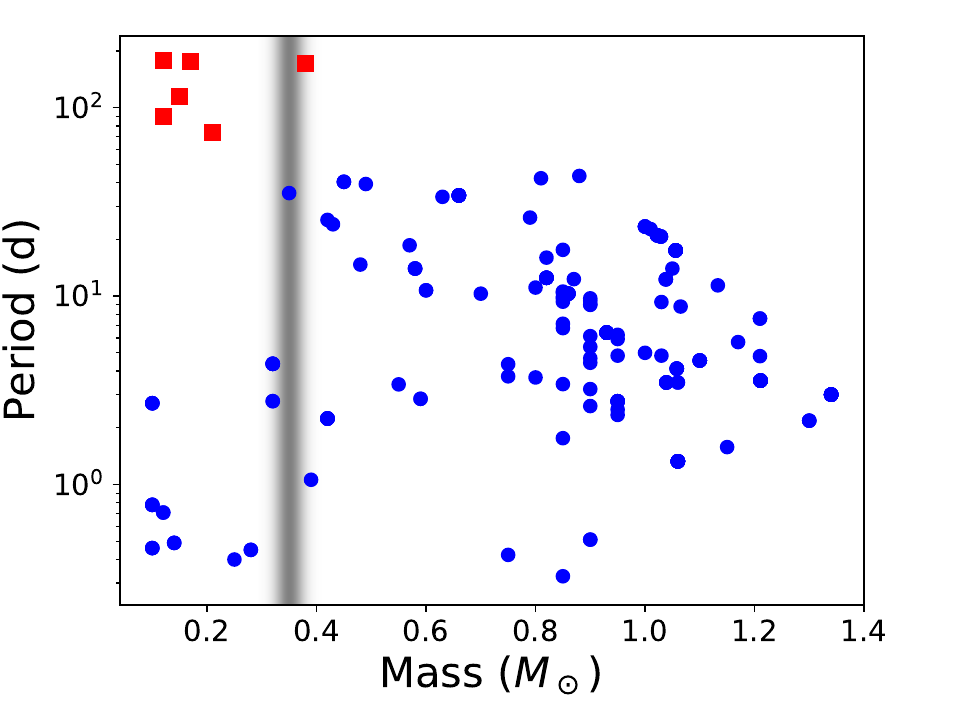}
    \caption{The rotation periods and masses of the sample of stars used in this study. A set of slowly rotating M dwarfs are shown with red square symbols which we will pay special attention to in the rest of this work. The vertical gray band represents the approximate location of the fully convective boundary.}
    \label{fig:Params}
\end{figure}

\section{Sample details}
\label{sec:Sample}
\subsection{Sample properties}
\label{subsec:SamProp}
For this study, we compile a sample of stars that have had their large-scale photospheric magnetic fields reconstructed with the ZDI technique. These stars have been observed over the last several decades and the ZDI maps of their large-scale magnetic fields have been published in the following papers: \cite{Donati2003,Donati2008,Petit2008,Morin2008,Morin2008b,Morin2010,Marsden2011,Jeffers2014,Waite2015,Mengel2016,Hebrard2016,doNascimento2016,Fares2009,Fares2010,Fares2012,Fares2013,Fares2017,Waite2017,BoroSaikia2015,BoroSaikia2016,BoroSaikia2018,Folsom2016,Folsom2018,Folsom2018b,Klein2021,Brown2021,Brown2024,Bellotti2023,Bellotti2024,Bellotti2024b,Lehmann2024}; Smith et al. (in prep). Many of the stars in our sample have been observed repeatedly over numerous epochs resulting in multiple ZDI maps for those stars. In total, our sample contains 260 ZDI maps of 96 stars. For the calculations in this work, we require various physical properties of the stars in our sample. These parameters are listed in table \ref{tab:Params} (as well as other parameters derived throughout this work) and their values are taken from the original paper in which each ZDI map was published. Where parameters were not available in these publications, they were taken from \citet{Reiners2007}, \citet{Strassmeier2009}, \citet{Lepine2013}, \citet{Marsden2014} and \citet{Pineda2021}. Figure \ref{fig:Params} shows the rotation periods and masses of the stars in our sample with the vertical dashed band indicating the approximate location of the FC boundary. Throughout this work, we will pay special attention to the sub-sample of stars identified with red squares in fig. \ref{fig:Params}. These are the fully convective M dwarfs with extremely long rotation periods mapped by \citet{Klein2021} and \citet{Lehmann2024} that occupy an area of parameter space that, until recently, had not been probed with ZDI. We note that one of these stars, Gl 408 lies to the right of the FC bounadry suggesting that it is a PC star. However, as we will show in section \ref{sec:MagProp}, all the stars identified with red squares, including Gl 408, have very similar magnetic properties. As such, it makes sense to group them together in our analysis. We will discuss this point further in section \ref{subsec:Gl408}.

\begin{table}
	\centering
	\caption{A summary of the details of each star/ZDI map in our sample. The full electronic table containing these values for our sample can be found online.}
	\begin{tabular}{lcc}
		\hline
		& Units & Description\\
        \hline
        Name & - & Name or ID of each star\\
        Reference & - & Citation to the paper the ZDI\\
        & & map was published in\\
        $M_\star$ & $M_\odot$ & Stellar mass\\
        $R_\star$ & $R_\odot$ & Stellar radius\\
        $T_{\rm eff}$ & K & Stellar effective temperature\\
        $P_{\rm rot}$ & days & Stellar rotation period\\
        $\B{}$ & G & Average field strength from ZDI\\
        $\B{dip}$ & G & Average dipole field strength from ZDI\\
        $\B{quad}$ & G & Average quadrupole field strength from ZDI\\
        $\B{oct}$ & G & Average octupole field strength from ZDI\\
        $\BE{pol}$ & G$^2$ & Average energy density in\\
        & & poloidal modes from ZDI\\
        $\BE{tor}$ & G$^2$ & Average energy density in\\
        & & toroidal modes from ZDI\\
        $\BE{axi}$ & G$^2$ & Average energy density in\\
        & & axisymmetric modes from ZDI\\
        $\frac{T_{\rm pred}}{T_\star}\big\rvert_{\rm d}$ & - & Torque ratio assuming the dipole is dominant\\
        ${\rm min} \left( \frac{T_{\rm pred}}{T_\star} \right)$ & - & Minimum value of the torque ratio\\
        ${\rm max} \left( \frac{T_{\rm pred}}{T_\star} \right)$ & - & Maximum value of the torque ratio\\
		\hline
	\end{tabular}
 \label{tab:Params}
\end{table}

\subsection{Convective turnover times}
\label{subsec:TurnoverTimes}
One of the key parameters we require in this work is the convective turnover time. There are a number of different definitions but, broadly speaking, they all describe the characteristic time-scale on which convective motions occur in stellar interiors. The turnover time is important since it, along with the rotation period, defines a star's Rossby number. Many different methods and prescriptions exist in the literature for estimating turnover times \citep[e.g.][]{Barnes2010,Corsaro2021,Landin2023,Chiti2024,Metcalfe2024}. In \citet{See2019ZBvsZDI} and \citet{See2019Geom}, we previously used the prescription presented by \citet{Cranmer2011}. However, this prescription is only valid for stars with effective temperatures hotter than $\sim$3300K. Since many of the stars we are focusing on in this work are cooler than this, we need a new prescription.

\begin{figure}
	\includegraphics[trim=0mm 10mm 0mm 0mm,width=\columnwidth]{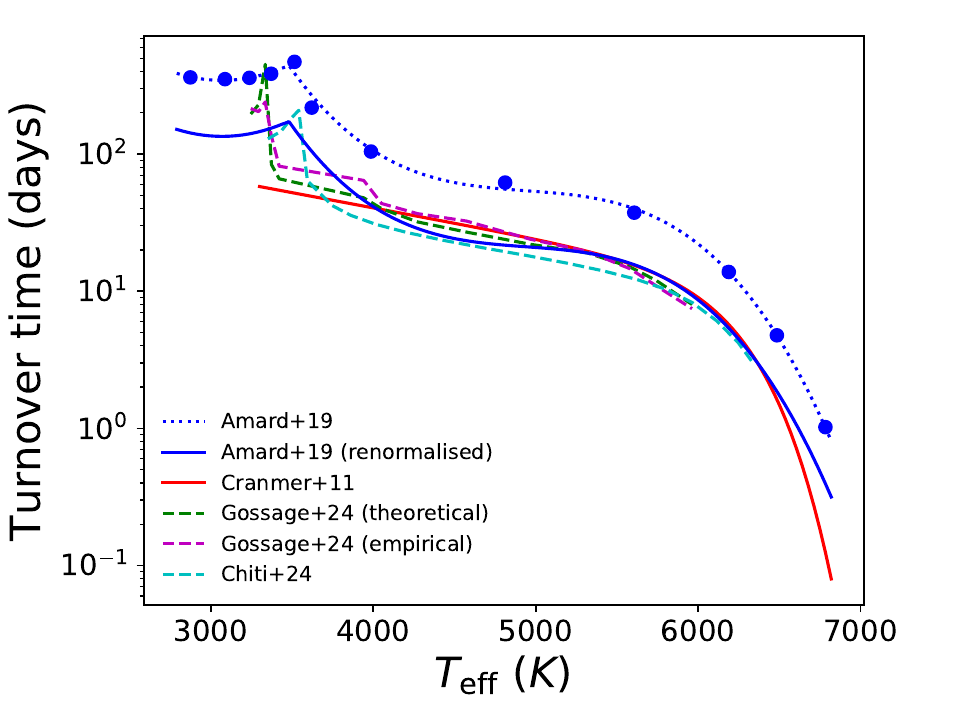}
    \caption{Convective turnover time as a function of effective temperature. The blue circular points are the global turnover times taken from the the individual models of \citet{Amard2019}. The blue dotted curve is a fit to these data points. The red curve shows the turnover time prescription of \citet{Cranmer2011}. The solid blue curve is the dotted blue curve normalised, such that it has the same turnover time value as the prescription of \citet{Cranmer2011} at the solar effective temperature. The offset between the blue dotted curve and the red curve is due to where the turnover times are calculated for these two prescriptions (see section \ref{subsec:TurnoverTimes}).} The analytic expression of the solid blue curve is given by equation (\ref{eq:TauPres}). The remaining dashed curves are prescriptions from \citet{Gossage2024} and \citet{Chiti2024}.
    \label{fig:Tau}
\end{figure}

In this work, we present a new prescription that is constructed using a set of structure models with masses in the range $0.1M_\odot$ to $1.4M_\odot$. These models were computed based on the physics described in \citet{Amard2019}. From each model, we take the effective temperature and global turnover time at the zero age main sequence. These are plotted in fig. \ref{fig:Tau} with blue circles. The global turnover time is calculated as

\begin{equation}
    \tau_{\rm G} = \int_{r_{\rm BCE}}^{R_\star} \frac{dr}{v_{\rm c}(r)},
    \label{eq:TauDef}
\end{equation}
where $r_{\rm BCE}$ is the base of the convective envelope, $R_\star$ is the stellar radius and $v_{\rm c}(r)$ is the local convective velocity. The global turnover time describes how long it takes a parcel of fluid to traverse the width of the convective envelope.

Figure \ref{fig:Tau} shows that the global turnover time increases with decreasing temperature until a maxima is reached at $\sim$3500K near the fully convective boundary. We fit an analytic function to these points consisting of a second order polynomial to data points from the five coolest models and a third order polynomial to the data points from the eight hottest models. This fit is shown with the dotted blue line in fig. \ref{fig:Tau}.

As can be clearly seen, the global turnover times derived from the models of \citet{Amard2019} are significantly larger than the ones from the prescription of \citet[][solid red line]{Cranmer2011}. When constructing their prescription, \citet{Cranmer2011} used the structure models of \citet{Gunn1998} who calculated local convective turnover times at one pressure scale height above the base of the convective zone. However, this approach to measuring turnover times for the coolest stars is not possible since their pressure scale height is larger than the stellar radius which is why we resort to calculating global turnover times instead (see also \citet{Somers2017} for alternative methods of calculating M dwarf turnover times).

Although there is no inherent problem with using global over local turnover times, it is still desirable to have our prescription provide turnover times that are roughly comparable with previous prescriptions. To do this, we introduce a scaling factor of 0.39. This value is chosen such that our prescription predicts the same turnover time as the prescription of \citet{Cranmer2011} at the solar temperature. We note that the introduction of this scaling factor does not affect the conclusions of this paper. The effect of this scaling factor is simply to shift all the data points by a fixed factor along the x axis on any plots involving the Rossby number. Our final prescription, that we use throughout the rest of this work, is shown as the solid blue line in fig. \ref{fig:Tau} and is given by

\begin{equation}
    \log \tau = {\rm min} \begin{cases}
            & \log 0.39 + a_1 T_{\rm eff}^2 + a_2 T_{\rm eff} + a_3 \\
            & \log 0.39 + b_1 T_{\rm eff}^3 + b_2 T_{\rm eff}^2 + b_3 T_{\rm eff} + b_4
    \end{cases}
    \label{eq:TauPres}
\end{equation}
where the values of the fit coefficients can be found in table \ref{tab:TauPres}. Looking at equation (\ref{eq:TauPres}) in fig. \ref{fig:Tau}, we see that there is a maximum in the turnover times at around $\sim$3500 K which is around the location of the fully convective boundary. The turnover times then steeply drop off towards hotter effective temperatures. Such a feature can be seen in other turnover time prescriptions derived from stellar structure models \citep[e.g.][]{Barnes2010,Chiti2024} as well as recent empirical turnover time prescriptions \citep{Gossage2024}. The peak occurs because, at this effective temperature, the convective envelope size is maximised (cooler stars are smaller which limits the size of the convection zone while more and more of the stellar interior is taken up by the radiative core in hotter stars).

\begin{table}
	\centering
	\caption{Best fit coefficients for equation (\ref{eq:TauPres})}
	\begin{tabular}{lc}
		\hline
		Coefficient & Value\\
        \hline
        $a_1$ & $6.52112823 \times 10^{-7}$\\
        $a_2$ & $-4.00355099 \times 10^{-3}$\\
        $a_3$ & $8.68234621$\\
        $b_1$ & $-2.51904051 \times 10^{-10}$\\
        $b_2$ & $3.73613409 \times 10^{-6}$\\
        $b_3$ & $-1.85566042 \times 10^{-2}$\\
        $b_4$ & $32.5950535$\\
		\hline
	\end{tabular}
 \label{tab:TauPres}
\end{table}

\section{Magnetic properties}
\label{sec:MagProp}
In this section, we investigate the magnetic properties of our sample. When reconstructing the magnetic fields of stars with ZDI, it is typical to represent the final map as a superposition of spherical harmonic modes. For this work, we follow the convention outlined in appendix B of \citet{Folsom2018}, which we reproduce here:

\begin{equation}
    B_{\rm r}(\theta,\phi) = \sum_{lm} {\rm Re}[\alpha_{lm} Y_{lm}(\theta,\phi)],
\label{eq:Br}
\end{equation}

\begin{equation}
    B_{\rm \theta}(\theta,\phi) = -\sum_{lm} {\rm Re}[\beta_{lm} Z_{lm}(\theta,\phi) + \gamma_{lm}X_{lm}(\theta,\phi)],
\label{eq:Bt}
\end{equation}

\begin{equation}
    B_{\rm \phi}(\theta,\phi) = -\sum_{lm} {\rm Re}[\beta_{lm} X_{lm}(\theta,\phi) - \gamma_{lm}Z_{lm}(\theta,\phi)],
\label{eq:Bp}
\end{equation}
where

\begin{equation}
    Y_{lm}(\theta,\phi) = c_{lm} P_{lm}(\cos\theta)e^{im\phi},
\label{eq:Ylm}
\end{equation}

\begin{equation}
    X_{lm}(\theta,\phi) = \frac{c_{lm}}{l+1} \frac{im}{\sin\theta} P_{lm}(\cos\theta)e^{im\phi},
\label{eq:Xlm}
\end{equation}

\begin{equation}
    Z_{lm}(\theta,\phi) = \frac{c_{lm}}{l+1} \frac{\partial P_{lm}(\cos\theta)}{\partial \theta} e^{im\phi},
\label{eq:Zlm}
\end{equation}
and

\begin{equation}
    c_{lm} = \sqrt{\frac{2l+1}{4\pi} \frac{(l-m)!}{(l+m)!}}.
\label{eq:clm}
\end{equation}
In these equations, $r$, $\theta$ and $\phi$ are the radial, meridional and azimuthal coordinates, $P_{lm}(\cos\theta)$ are the associated Legendre polynomials with the subscripts $l$ and $m$ indicating the degree and order of the polynomial. In this convention, the ZDI map is defined by the $\alpha_{lm}$, $\beta_{lm}$ and $\gamma_{lm}$ coefficients which are determined by fitting to a time series of spectropolarimetric observations. The different magnetic properties that we will analyse in this section are defined by different subsets of these coefficients. Before moving onto the results, it is worth briefly discussing uncertainties. Uncertainties for the magnetic field strengths derived from ZDI maps are not always calculated/reported in the original publications. However, when they are reported, they are typically a few tens of percent. These uncertainties are usually derived by determining how much a given ZDI map changes when the input parameters used to construct it, e.g. rotation period, are varied within their uncertainties.

\begin{figure}
    \includegraphics[trim=0mm 10mm 0mm 0mm,width=\columnwidth]{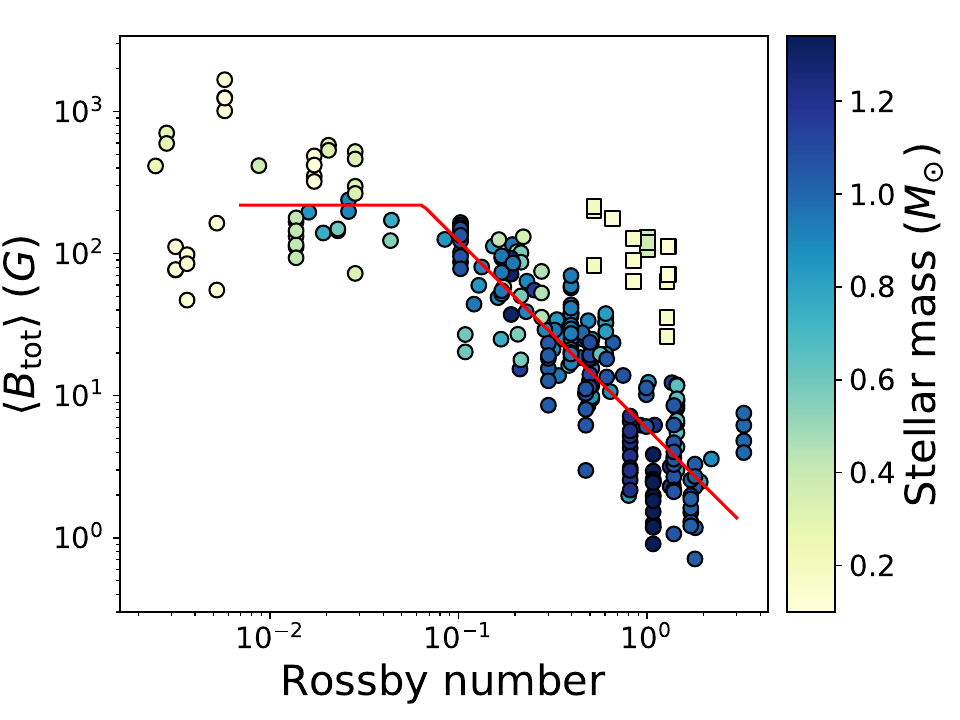}
    \caption{The average unsigned magnetic field strength of our sample of stars vs Rossby number. The slowly rotating M dwarfs identified in fig. \ref{fig:Params} are shown with square symbols. The red curve is a fit using equation \ref{eq:ActRotRel}. The slowly rotating M dwarfs and stars with Rossby numbers smaller than 0.006 are not included in this fit (see text for further details).}
    \label{fig:BVsRo}
\end{figure}

\begin{figure*}
    \includegraphics[trim=0mm 10mm 0mm 0mm,width=\textwidth]{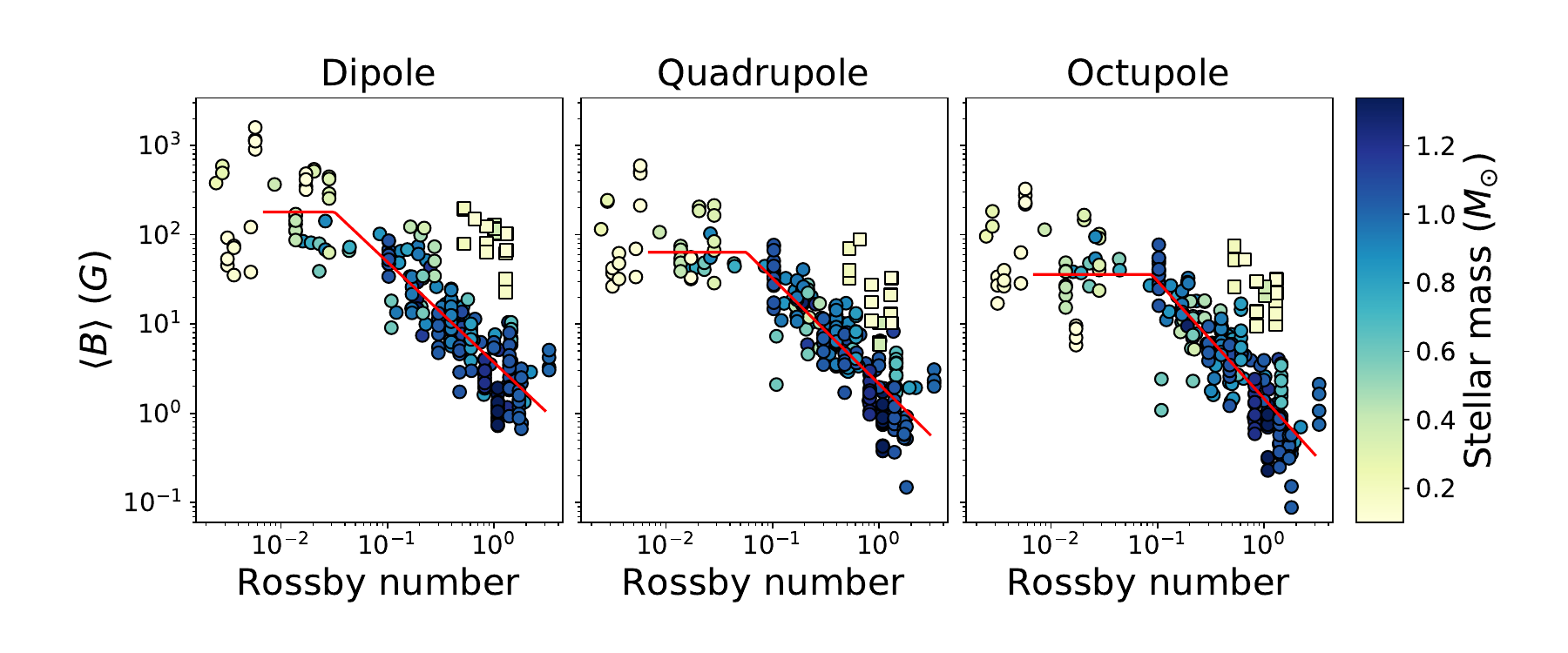}
    \caption{The average unsigned dipolar, quadrupolar and octupolar magnetic field strengths of our sample of stars vs Rossby number. The slowly rotating M dwarfs identified in fig. \ref{fig:Params} are shown with square symbols. The red curve is a fit using equation \ref{eq:ActRotRel}. The slowly rotating M dwarfs and stars with Rossby numbers smaller than 0.006 are not included in this fit (see text for further details).}
    \label{fig:DQOVsRo}
\end{figure*}

\subsection{Magnetic field - Rossby number relations}
\label{subsec:ActRotRel}
As noted in the introduction, it is thought that the generation of magnetic fields via dynamo action can be parameterised by the Rossby number. Figure \ref{fig:BVsRo} shows the average magnetic field strengths of each ZDI map from our sample as a function of Rossby number. Vertical ``lines'' of data points represent individual stars that have been mapped over multiple epochs with the spread of field strengths being due to the temporal evolution of the star. Similar to previous ZDI studies \citep[e.g.][]{Vidotto2014,Folsom2018,See2019ZBvsZDI}, the majority of the stars follows the well known activity-rotation relation and separate into saturated and unsaturated regimes. We also confirm that the slowly rotating FC M dwarfs, i.e. those coloured red in fig. \ref{fig:Params} and shown with square symbols in fig. \ref{fig:BVsRo}, lie systematically above the trend followed by the rest of the stars. This result was first noted by \citet{Lehmann2024}. 

As well as looking at the total magnetic field strength of our stars, we can also look at the field strengths of the poloidal dipolar, quadrupolar and octupolar components. These components are important as they are the ones that determine the strength of magnetic braking from stellar winds \citep[see e.g.][and section \ref{sec:Spindown}]{See2019Geom}. In terms of the spherical harmonic decomposition, these components are given by the $\alpha_{lm}$ \& $\beta_{lm}$ coefficients where the $l$ mode can take the values 1, 2 or 3. Figure \ref{fig:DQOVsRo} shows how these components vary with Rossby number. We see that the dipolar, quadrupolar and octupolar components follow the same qualitative behaviour as the total magnetic field strength shown in fig. \ref{fig:BVsRo}, i.e. the majority of stars separate out into the saturated and unsaturated regimes while the slowly rotating FC M dwarfs lie above the main bulk of the stars. 


\begin{table}
	\centering
	\caption{The best fit values for equation (\ref{eq:ActRotRel}) when considering different components of the magnetic field strength.}
	\label{tab:ActRotRelFitVals}
	\begin{tabular}{lccc}
		\hline
		  & $B_{\rm sat}$ (G) & $\rm Ro_{crit}$ & $\beta$\\
		\hline
Total &  219 $\pm$ 29 &  0.065 $\pm$ 0.009 &  -1.32 $\pm$ 0.05 \\
Dipole &  179 $\pm$ 26 &  0.032 $\pm$ 0.006 &  -1.13 $\pm$ 0.06 \\
Quadrupole &  64 $\pm$ 8 &  0.057 $\pm$ 0.009 &  -1.19 $\pm$ 0.05 \\
Octupole &  36 $\pm$ 5 &  0.088 $\pm$ 0.013 &  -1.32 $\pm$ 0.06 \\
		\hline
	\end{tabular}
\end{table}

Although it is clear that there is not a simple dependence on the Rossby number in figs. \ref{fig:BVsRo} and \ref{fig:DQOVsRo}, we can still fit an activity-rotation relation. We use a fit of the form

\begin{equation}
\begin{split}
    \langle B \rangle & = B_{\rm sat} &{\rm for\ Ro < Ro_{crit}} \\
    \langle B \rangle & = B_{\rm sat} {\rm \left( \frac{Ro}{Ro_{crit}} \right)^\beta} &{\rm for\ Ro \geq Ro_{crit}},
\end{split}
\label{eq:ActRotRel}
\end{equation}
where $\langle B \rangle$ is the field strength recovered from ZDI averaged over the stellar surface, $\rm Ro$ is the Rossby number, $B_{\rm sat}$ is the field strength in the saturated regime, $\rm Ro_{crit}$ is the critical Rossby number separating the saturated and unsaturated regimes and $\beta$ is the power-law index or slope of the unsaturated regime. Similarly to \citet{See2019ZBvsZDI} and \citet{See2019Geom}, when fitting equation (\ref{eq:ActRotRel}), we exclude stars with $\rm Ro < 0.006$. These are the stars with bimodal magnetic fields noted by \citet{Morin2010} and it is unclear how they fit into the saturated-unsaturated paradigm. Additionally, we also exclude the slowly rotating M dwarfs (the ones shown as squares) from the fit as they clearly do not follow the same trend. The best fit parameters for the total, dipolar, quadrupole and octupolar field strengths are shown in table \ref{tab:ActRotRelFitVals} and shown on figs. \ref{fig:BVsRo} and \ref{fig:DQOVsRo} with red lines.

\begin{figure}
	\includegraphics[trim=0mm 10mm 0mm 0mm,width=\columnwidth]{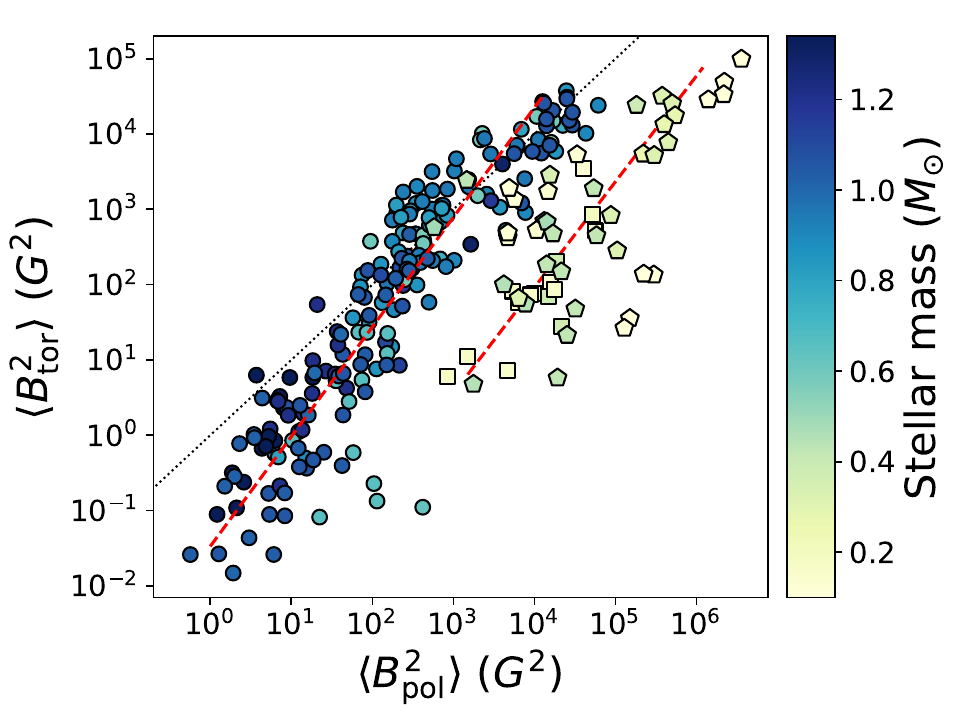}
    \caption{Toroidal magnetic energy density vs poloidal magnetic energy density for our sample of stars coloured by stellar mass. Circle symbols indicate stars more massive than 0.5$M_\odot$, pentagon symbols indicate stars less massive than 0.5$M_\odot$ and square symbols indicate stars belonging to the set of slowly rotating M dwarfs identified in fig. \ref{fig:Params}. The red dashed lines are fits to stars above 0.5$M_\odot$ and stars below 0.5$M_\odot$. The black dotted line shows $\langle B_{\rm tor}^2\rangle = \langle B_{\rm pol}^2\rangle$.}
    \label{fig:TorVsPol}
\end{figure}

\begin{figure*}
	\includegraphics[trim=0mm 10mm 0mm 0mm,width=0.8\textwidth]{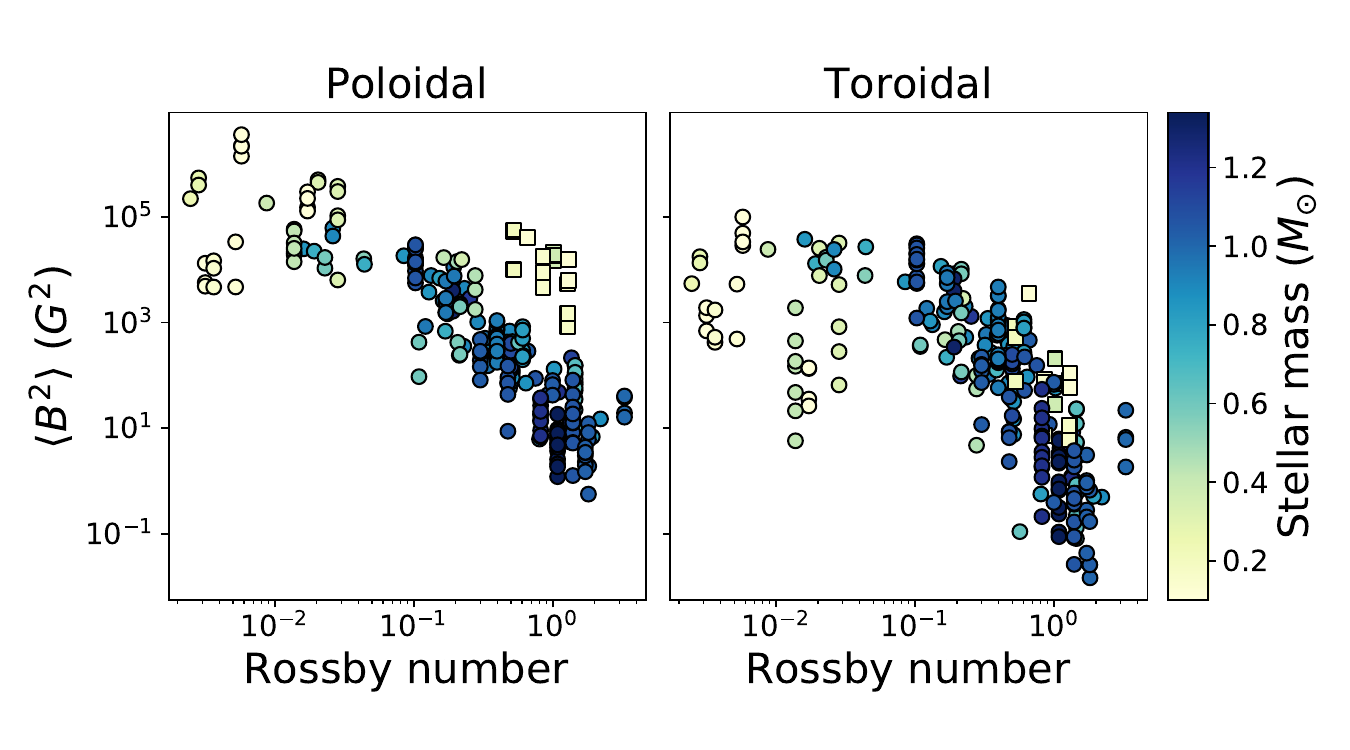}
    \caption{Poloidal (left panel) and toroidal (right panel) magnetic energy density vs Rossby number for our sample of stars coloured by stellar mass. The slowly rotating M dwarfs identified in fig. \ref{fig:Params} are shown with square symbols.}
    \label{fig:PolTorVsRo}
\end{figure*}

\begin{figure}
    \includegraphics[trim=0mm 10mm 0mm 0mm,width=\columnwidth]{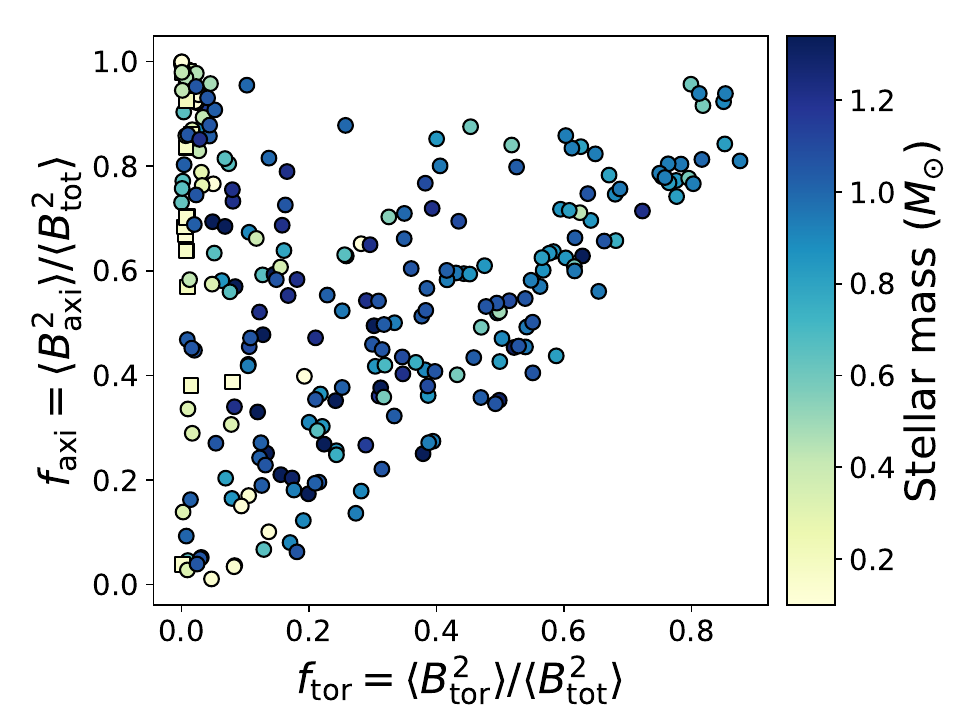}
    \caption{The fraction of magnetic energy contained in axisymmetric modes vs the fraction of magnetic energy contained in the toroidal field for our sample of stars coloured by stellar mass. The slowly rotating M dwarfs identified in fig. \ref{fig:Params} are shown with square symbols.}
    \label{fig:AxiVsTor}
\end{figure}

\subsection{Poloidal and toroidal magnetic fields}
\label{subsec:PolTor}
When studying stellar magnetic fields, it is common to split the field up into its poloidal and toroidal components. The poloidal component of the field is represented by the $\alpha_{lm}$ and $\beta_{lm}$ coefficients in equations (\ref{eq:Br})-(\ref{eq:Bp}) while the toroidal component is represented by the $\gamma_{lm}$ coefficients. In fig. \ref{fig:TorVsPol}, we show the toroidal magnetic energy density, $\langle B_{\rm tor}^2\rangle$, against the poloidal magnetic energy density, $\langle B_{\rm pol}^2\rangle$, for our sample of stars. Overall, we see a positive correlation between the poloidal and toroidal magnetic field energy densities with the majority of our sample being poloidal dominated. 

We also previously studied the relationship between the toroidal and poloidal magnetic energies in \citet{See2015} based on a smaller sample of ZDI maps (90 maps in that work vs 260 maps in this work). In that work, we suggested that stars lie on two sequences, one formed by stars less massive than 0.5$M_\odot$ and one formed by stars more massive than 0.5$M_\odot$. Two sequences are also evident in fig. \ref{fig:TorVsPol} with our expanded sample of stars, one composed of higher mass stars and the other composed of lower mass stars. Similarly to \citet{See2015}, we use 0.5$M_\odot$ as the stellar mass that separates the two sequences \citep[see also][]{Donati2009}. However, we note that the exact mass that separates the two sequences is not clear cut. For the remainder of this work, we will use 0.5$M_\odot$ as the dividing mass when discussing these two sequences but we also explore the effect of using other masses in appendix \ref{sec:MassSplit}. In fig. \ref{fig:TorVsPol} we see that $<0.5M_\odot$ stars generally have smaller toroidal magnetic energies than the $>0.5M_\odot$ stars for a given poloidal magnetic energy. Both sequences have some scatter associated with them which can be attributed to magnetic cycles or general variability. Using a orthogonal distance regression we fit power laws of the form $\log_{10} \langle B_{\rm tor}^2\rangle = {\rm m} \log_{10} \langle B_{\rm pol}^2\rangle + {\rm c}$ to the two sequences which are shown with red dashed lines. We find fit values of $\rm m=1.46 \pm 0.04$, $\rm c= -1.47\pm 0.11$ for the stars more massive than $0.5M_\odot$ (circle points) and $\rm m=1.40 \pm 0.17$, $\rm c=-3.65 \pm 0.79$ for the stars less massive than $0.5M_\odot$ (square and pentagon points). Our expanded sample of stars offers some new insights into the toroidal and poloidal energies of low-mass stars compared to \citet{See2015}. Firstly, we see that the two sequences are almost parallel in fig. \ref{fig:TorVsPol} whereas in \citet{See2015} the slope of the $<0.5M_\odot$ stars was much shallower than the slope of the $>0.5M_\odot$ stars. We also see that there appears to be more scatter in the $<0.5M_\odot$ sequence than the $>0.5M_\odot$ sequence. Finally, we see a number of data points in the $>0.5M_\odot$ sequence at $\langle B_{\rm pol}^2\rangle \sim 100{\rm G^2}$ and $\langle B_{\rm tor}^2\rangle \sim 0.15{\rm G^2}$ that appear to be outliers when compared with the rest of the sequence. These data points belong to the stars 61 Cyg A (observed over multiple epochs \citep{BoroSaikia2016,BoroSaikia2018}) and GJ 205 \citep{Hebrard2016}. Interestingly, these stars are very similar with masses of 0.66$M_\odot$ and 0.63$M_\odot$ and rotation periods of 34.2 days and 33.6 days respectively. It is unclear why these data points lie further away from the sequence formed by the rest of the $>0.5M_\odot$ stars. We also note that, of the 12 data points in fig. \ref{fig:TorVsPol} associated with 61 Cyg A, only 3 or 4 are in this group of outliers with the remaining data points lying within the the main trend of the $>0.5M_\odot$ stars. Therefore, it seems that 61 Cyg A only spends part of its activity cycle off the main trend.

Similarly to section \ref{subsec:ActRotRel}, we can look at how the poloidal and toroidal magnetic energies vary with Rossby number. This is shown in fig. \ref{fig:PolTorVsRo}. Broadly speaking, the poloidal magnetic energy densities follows the same trends identified for the various field components in section \ref{subsec:ActRotRel}. That is, the poloidal magnetic energy densities split into saturated and unsaturated regimes. As before, the slowly rotating M dwarfs lie significantly above their partially convective counterparts in the unsaturated regime. However, the situation is slightly different for the toroidal energy densities. In the right panel of fig. \ref{fig:PolTorVsRo}, the slowly rotating M dwarfs follow the same trend as their partially convective counterparts. Additionally, in the saturated regime, many of the $<0.5M_\odot$ stars have much smaller toroidal magnetic energy densities than the notional saturation level. This behavior is a natural consequence of the trends seen in fig. \ref{fig:TorVsPol}. That figure showed that, for comparable values of $\BE{tor}$, $<0.5M_\odot$ stars have larger $\BE{pol}$ values than $>0.5M_\odot$ stars, thus resulting in the $<0.5M_\odot$ stars being lower down, relative to the $>0.5M_\odot$ stars, in the $\BE{tor}$ vs Rossby number diagram than they are in the $\BE{pol}$ vs Rossby number diagram.

In \citet{See2015}, we also looked at how axisymmetric the magnetic fields of low-mass stars are as a function of the toroidal energy fraction. In both that work and this one, we consider axisymmetric field modes to be those with $\rm m=0$ in the spherical harmonic decomposition. We found that stars with a high fraction of their magnetic energy in the poloidal component have a large spread in how axisymmetric their magnetic fields are. In contrast, toroidal magnetic fields tend to be very axisymmetric and, as such, stars with dominantly toroidal fields are very axiysmmetric overall. We find that this is still the case even with our expanded sample as shown in fig. \ref{fig:AxiVsTor}.

\section{Stellar torque}
\label{sec:Spindown}
In this section, we investigate the rate at which stars in our sample are losing angular momentum. To do this we make use of a so called `braking law'. These are analytic expressions used to estimate the angular momentum loss rate, or torque, of a star as a function of its properties and are constructed using magnetohydrodynamic simulations \citep{Matt2012,Reville2015,Finley2017,Pantolmos2017}. In this study, we use the braking law of \citet{Finley2018} which is a twice broken power law given by

\begin{equation}
    T_\star = \dot{M} \Omega_\star \langle R_{\rm A} \rangle^2,
\label{eq:Torque}
\end{equation}
where $\dot{M}$ is the stellar mass-loss rate, $\Omega_\star$ is the rotation rate and $\langle R_{\rm A} \rangle$ is the torque averaged Alfv\'en radius which is given by

\begin{subnumcases}{\langle R_{\rm A} \rangle / R_\star = {\rm max} \label{eq:rA}}
    K_{\rm d} [ \mathcal{R}_{\rm d}^2 \Upsilon ] ^{m_{\rm d}}\label{eq:rADip} \\
    K_{\rm q} [ ( \mathcal{R}_{\rm d} + \mathcal{R}_{\rm q} )^2 \Upsilon ]^{m_{\rm q}} \label{eq:rAQuad} \\
    K_{\rm o} [ (\mathcal{R}_{\rm d} + \mathcal{R}_{\rm q} + \mathcal{R}_{\rm o})^2 \Upsilon ]^{m_{\rm o}}. \label{eq:rAOct}
\end{subnumcases}
Here $R_\star$ is the radius of the star, $\Upsilon=\frac{B_\star^2 R_\star^2}{\dot{M} v_{\rm esc}}$ is the wind magnetization parameter, and $v_{\rm esc}$ is the stellar escape velocity. The field ratios of the dipole, quadrupole and octupole components, $\mathcal{R}_{\rm d}$, $\mathcal{R}_{\rm q}$ and $\mathcal{R}_{\rm o}$ are given by $B_{\rm d}/B_\star$, $B_{\rm q}/B_\star$ and $B_{\rm o}/B_\star$ respectively with $B_\star = B_{\rm d}+B_{\rm q}+B_{\rm o}$. Lastly, $K_{\rm d}=1.53$, $K_{\rm q}=1.7$, $K_{\rm o}=1.8$, $m_{\rm d}=0.229$, $m_{\rm q}=0.134$ and $m_{\rm o}=0.087$, are fit parameters derived from the MHD simulations of \citet{Finley2018}. It is worth noting that equation (\ref{eq:rADip}) only depends on the dipole field strength while equations (\ref{eq:rAQuad}) and (\ref{eq:rAOct}) also depend on the quadrupole and octupole field strength. As we showed in \citet{See2019Geom}, the mass-loss rate is the parameter that controls which of these equations is maximal and, hence, which equation determines the torque.

When constructing equation (\ref{eq:rA}), \citet{Finley2018} only considered axisymmetric field configurations. In their work, $B_{\rm d}$, $B_{\rm q}$ and $B_{\rm o}$ are the field strengths at the rotation pole or, equivalently, the magnetic pole of the star. However, the ZDI maps used in our work contains both axisymmetric and non-axisymmetric magnetic field components. Rather than using the field strength at the rotation pole, we will use the surface averaged unsigned field strengths shown in fig. \ref{fig:DQOVsRo} for $B_{\rm d}$, $B_{\rm q}$ and $B_{\rm o}$ in equation (\ref{eq:rA}), similarly to method in \citet{See2019Geom}. 

One final quantity is required to evaluate equation (\ref{eq:Torque}) for our sample of stars which is the mass-loss rate from the stellar wind. Due to the diffuse nature of the winds of low-mass stars, mass-loss rates are extremely difficult to measure. Although techniques exist to observationally constrain stellar mass-loss rates, they tend to either only provide upper limits or only be applicable under certain conditions \citep{Vidotto2017,Fichtinger2017,Jardine2019,Wood2021,Kislyakova2024}. As such, systematically estimating mass-loss rates for large samples of stars remains difficult. In \citet{See2019Geom}, we attempted to estimate mass-loss rates using numerical models \citep{Cranmer2011} or by inferring them from rotation evolution models \citep{Matt2015}. However, the slowly rotating M dwarfs in this work fall outside the range in which the model of \citet{Cranmer2011} is applicable while the rotation evolution model of \citet{Matt2015} makes the assumption that the field strengths of stars follow a single activity-rotation relation which we have shown is not true in section \ref{subsec:ActRotRel}. 

We will side step the problem of estimating mass-loss rates by calculating a ratio of torques, $T_{\rm pred}/T_\star$, for each star in our sample. Here $T_\star$ is the torque for each star calculated using the dipole, quadrupole and octupole field strengths presented in fig. \ref{fig:DQOVsRo} while $T_{\rm pred}$ is a predicted torque where the dipole, quadrupole and octupole field strengths are estimated using equation (\ref{eq:ActRotRel}) and the fit parameters in table \ref{tab:ActRotRelFitVals}. This torque ratio tells us how over- or under-estimated a torque calculation for each star would be if we did not know its dipole, quadrupole and octupole field strengths and, instead, assumed they followed a standard Rossby number scaling relation. Additionally, we can calculate the range of values this torque ratio can take without knowing the star's mass-loss rate as it cancels out in the ratio. This range of values is given by

\begin{equation}
\begin{split}
    {\rm min}\left( \frac{T_{\rm pred}}{T_\star} \right) &= {\rm min} \left\{ \frac{T_{\rm pred}}{T_\star}\bigg\rvert_{\rm d}, \frac{T_{\rm pred}}{T_\star}\bigg\rvert_{\rm q}, \frac{T_{\rm pred}}{T_\star}\bigg\rvert_{\rm o} \right\}, \\
    {\rm max}\left( \frac{T_{\rm pred}}{T_\star} \right) &= {\rm max} \left\{ \frac{T_{\rm pred}}{T_\star}\bigg\rvert_{\rm d}, \frac{T_{\rm pred}}{T_\star}\bigg\rvert_{\rm q}, \frac{T_{\rm pred}}{T_\star}\bigg\rvert_{\rm o} \right\},
    \label{eq:TorqueRatioRange} 
\end{split}
\end{equation}
where
\begin{subequations}
\begin{align}
    \frac{T_{\rm pred}}{T_\star}\bigg\rvert_{\rm d} & = \left( \frac{\B{pred,d}}{\B{ZDI,d}} \right)^{4m_{\rm d}}, \label{eq:TorqueRatioDip} \\
    \frac{T_{\rm pred}}{T_\star}\bigg\rvert_{\rm q} & = \left( \frac{\B{pred,d} + \B{pred,q}}{\B{ZDI,d} + \B{ZDI,q}} \right)^{4m_{\rm q}}, \label{eq:TorqueRatioQuad} \\
    \frac{T_{\rm pred}}{T_\star}\bigg\rvert_{\rm o} & = \left( \frac{\B{pred,d} + \B{pred,q} + \B{pred,o}}{\B{ZDI,d} + \B{ZDI,q} + \B{ZDI,o}} \right)^{4m_{\rm o}}, \label{eq:TorqueRatioOct}
\end{align}
\label{eq:TorqueRatioComps}
\end{subequations}
i.e. the maxima and minima of the torque ratio is given by whichever of equations (\ref{eq:TorqueRatioDip}), (\ref{eq:TorqueRatioQuad}) or (\ref{eq:TorqueRatioOct}) is largest and smallest respectively. Before calculating the torque ratios for our sample, it is worth briefly explaining the form of equations (\ref{eq:TorqueRatioRange}) and (\ref{eq:TorqueRatioComps}). It is most simple to consider the case where the star's spin-down torque is dominated by the dipole component of the magnetic field first. A number of studies have shown that this is generally true for most stars \citep{Jardine2017,See2017,See2018,See2020} and that higher order modes like the quadrupole or octupole only need to be accounted for once the star's mass-loss rate exceeds some critical value \citep{See2019Geom}. When only considering the magnetic dipole, the braking law reduces down to just equations (\ref{eq:Torque}) and (\ref{eq:rADip}). The torque ratio is then simply given by equation (\ref{eq:TorqueRatioDip}) and is only a function of the predicted dipole field strength of the star and the dipole field strength recovered by ZDI. For the small number of cases where the dipole dominated approximation breaks down, the contribution of higher order field modes to the torque needs to be accounted for, i.e. the torque is no longer given by just equations (\ref{eq:Torque}) and (\ref{eq:rADip}) but also equations (\ref{eq:rAQuad}) and (\ref{eq:rAOct}). In this situation, it is no longer possible to exactly determine the torque ratio because we do not know the stellar mass-loss rate, and thus which regime of equation (\ref{eq:rA}) to use. However, it is possible to determine a lower and upper bound. Just as equation (\ref{eq:TorqueRatioDip}) can be derived from equation (\ref{eq:rADip}), equations (\ref{eq:TorqueRatioQuad}) and (\ref{eq:TorqueRatioOct}) can be derived from equations (\ref{eq:rAQuad}) and (\ref{eq:rAOct}) respectively. The true torque ratio falls somewhere between the maxima and minima of these three terms as stated by equation (\ref{eq:TorqueRatioRange}). The explanation for why the true torque ratio falls in this range is somewhat lengthy and so we present this in appendix \ref{sec:TorqueRatioExplain}.

\begin{figure}
    \includegraphics[trim=0mm 10mm 0mm 0mm,width=\columnwidth]{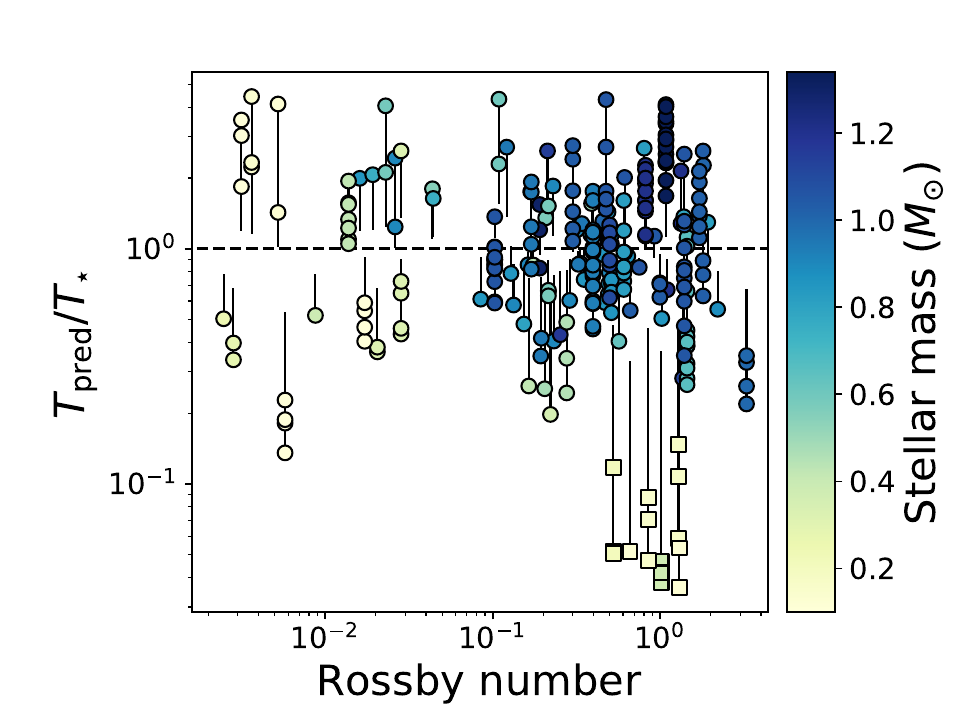}
    \caption{The torque ratio for our sample of stars vs Rossby number. $T_{\rm pred}$ is a predicted torque calculated using dipole, quadrupole and octupole field strengths estimated with equation (\ref{eq:ActRotRel}) and the fit parameters in table \ref{tab:ActRotRelFitVals} while $T_\star$ is calculated using the dipole, quadrupole and octupole field strengths presented in fig. \ref{fig:DQOVsRo}. The circle and square points show the torque ratio under the assumption that the torque is dominated by the dipole components while the full range of possible torque ratios for each star is shown by a vertical line.}
    \label{fig:TorqueRatio}
\end{figure}

Figure \ref{fig:TorqueRatio} shows the torque ratio against Rossby number for our sample of stars. The symbols represent the torque ratio under the assumption that the dipole component dominates the torques (equation (\ref{eq:TorqueRatioDip})) while vertical lines show the full range of possible torque ratios (equation (\ref{eq:TorqueRatioRange})). Looking at the dipole dominant case first, we see that the majority of our stars (circle points) lie around $T_{\rm pred}/T_\star=1$ with a scatter of roughly an order of magnitude. This scatter is mostly due to the fact that stellar magnetic fields are inherently time variable, e.g. due to activity cycles. In contrast to the rest of the stars, the slowly rotating M dwarfs (square points) have torque ratios of $\sim 0.1$ or less. Thus, if one did not know the true dipole field strengths of slowly rotating FC stars and used a standard Rossby number scaling, that is calibrated only to PC stars, to estimate the dipole field strength, one would underestimate their spin-down torque by an order of magnitude or more. 

Next, we consider the case where higher order field modes contribute to the spin-down torque, i.e. the vertical lines. The range of possible torque ratios for each star, when accounting for higher order field modes, generally lies closer to $T_{\rm pred}/T_\star=1$ than the dipole dominant case, i.e. the dipole dominant case generally represents the most extreme case of over- or under-estimating a stars true torque when using standard Rossby number scalings.

\section{Discussion}
\label{sec:Discussion}
\subsection{Regimes of dynamo operation}
\label{subsec:Dynamo}
The dynamo processes that are responsible for generating stellar magnetic fields are still not well understood. However, by studying the surface magnetic field properties of low-mass stars, as we have done in this work, we can place constraints on dynamo theory \citep[e.g.][]{Noraz2024}. In particular, abrupt changes in magnetic field properties, as a function of the physical properties of stars, could be an indication that different dynamo modes or processes operate in different physical regimes. In this work, we have seen two different instances of these types of abrupt changes in magnetic properties.

The first abrupt change of magnetic field properties occurs at $\sim 0.5 M_\odot$ which is clearly seen in the toroidal and poloidal magnetic energies (fig. \ref{fig:TorVsPol}). In this parameter space, stars more massive and less massive than 0.5$M_\odot$ form two distinct branches. An abrupt change of magnetic properties at $\sim 0.5M_\odot$ was also previously noted by \citet{Donati2008} and \citet{Gregory2012}. The existence of these two branches is difficult to explain. The physical stellar properties that one might expect to be relevant to the operation of a dynamo, e.g. the convective depth or the convective turnover time, all vary smoothly as a function of stellar mass at 0.5$M_\odot$, at least in 1D structure models. As such, one would naively expect that magnetic properties should also vary smoothly as a function of stellar mass at 0.5$M_\odot$ rather than the abrupt change we actually see. Adding to this puzzle is the lack of similar abrupt changes in behaviour at 0.5$M_\odot$ in other activity indicators.

The second abrupt change of magnetic field properties we analysed in this work occurs for slowly rotating FC M dwarfs. In general, it is expected that magnetic field strengths should be well parameterised by the Rossby number. However, we have seen in section \ref{subsec:ActRotRel} that the magnetic fields of the slowly rotating FC M dwarfs in our sample are stronger than expected in the unsaturated regime and do not follow the same Rossby number scaling as their PC counterparts \citep[see also][]{Lehmann2024}. This indicates that different dynamo processes are operating in FC and PC stars, at least in the unsaturated regime. Unlike the previous case, where it was unclear what was driving the change in magnetic properties at 0.5$M_\odot$, in this case, there seems to be a relatively obvious cause (although we will return to this point in section \ref{subsec:Gl408}). FC and PC stars have different physical structures (the lack, or presence, of a radiative core respectively) and this likely affects the way in which magnetic fields are generated in these two types of stars.

While the suggestion of different dynamo mechanisms in FC and PC stars may explain the magnetic field observations, it is possibly at odds with results from X-ray studies. Several studies have found that the X-ray emission of FC and PC stars both follow the same Rossby number scaling \citep{Wright2016,Wright2018,Gossage2024} contrary to the behaviour we have seen here and in \citet{Lehmann2024} for magnetic fields. This is potentially surprising since X-ray emission is a form of magnetic activity and so, one might expect that it should follow the same trends as magnetic fields. However, it should be noted that ZDI only probes the large-scale magnetic fields of a star \citep{Lehmann2019} while X-ray emission is a probe of small-scale magnetic fields. Therefore, it seems that there is a meaningful difference in the dynamos of FC and PC stars but only in a way that affects the generation of the large-scale magnetic fields.

Different dynamos operating in FC and PC stars is also consistent with the recent results of \citet{Lu2024}. These authors studied the rotation periods of FC and PC stars and showed that FC stars lose angular momentum more rapidly than PC stars, even when other parameters like mass or rotation rate are the same. Since angular momentum loss in low-mass stars is ultimately a magnetically driven phenomena, these authors suggested that the reason for the difference in angular momentum loss is that the dynamos of FC and PC stars operate in some fundamentally different way, similar to our conclusion in this work.

\subsection{A closer look at Gl 408}
\label{subsec:Gl408}
In section \ref{sec:Sample}, we identified a sub-sample of six stars (identified as the red squares in fig. \ref{fig:Params}), one of which is Gl 408. We then showed that the magnetic fields of the stars in this sub-sample do not follow the same Rossby number scaling as the rest of the stars we have studied (see section \ref{subsec:ActRotRel}). The majority of the stars in this sub-sample are FC and in section \ref{subsec:Dynamo}, we suggested that the reason the stars in this sub-sample follow a different Rossby number scaling is because different dynamo mechanisms are in operation in FC and PC stars. However, although Gl 408 is certainly close to being FC, there is a distinct possibility that it is actually PC, if only just. It has a mass of $0.38M_\odot$ which puts it on the PC side of the FC boundary (models typically predict the FC boundary occurs at roughly 0.35 $M_\odot$, e.g. \citet{Chabrier1997}). Additionally, it has a Gaia magnitude of $\rm M_G=9.831$. This places it on the PC side of the so-called Jao gap, an under-density in the Hertzsprung-Russell Diagram that is thought to coincide with the FC boundary \citep{Jao2018}. If Gl 408 is truly a PC star, it would cast doubt on our suggestion that different dynamo mechanisms operating in FC and PC stars can explain why the stars in our sub-sample follow a different Rossby number scaling to the rest of our sample.

We propose two possibilities to resolve this apparent discrepancy. The first is that Gl 408 could be FC despite the previously mentioned evidence. Determining whether a star is FC or PC can be challenging if that star is close to the FC boundary. One reason is because the FC boundary is not just a simple function of mass, but rather a complex function of both a star's effective temperature and metallicity \citep{Amard2019,Lu2024}. Additionally, different structure models make slightly different predictions for the location of the FC boundary. This problem is also complicated by the presence of the $^3$He instability in stars near the FC boundary which causes stars to periodically switch between FC and PC states early on in their evolution \citep{vanSaders2012,Baraffe2018}. Given these complicating factors, and the closeness of Gl 408 to the FC boundary, it is not outside the realm of possibility that Gl 408 could be FC, especially if we also take measurement errors into account.

However, if Gl 408 is truly PC, we would need to adjust our interpretation slightly. We would still suggest that different dynamo mechanisms are in operation for the six stars in our sub-sample compared to the other stars in our sample. However, rather than the transition between these two dynamo modes occurring exactly at the FC boundary, it would have to occur slightly before the onset of full convection. Given that Gl 408 is relatively close to the FC boundary, the shift of the dynamo transition location would not have to be large. Additionally, if Gl 408 is truly PC, this may inform us about the parts of the stellar interior involved in the generation of the stronger than expected magnetic fields in slowly rotating M dwarfs. It would suggest that, for these stars, the presence of a small deeply buried radiative core does not significantly affect the dynamo processes generating these fields compared to a star with a FC interior, hinting that the dynamo processes are mostly occurring slightly higher up in these stars and not near the core.

Another complicating factor is that Gl 408 has multiple conflicting rotation periods reported in the literature. \citet{Donati2023} and \citet{Lehmann2024} find rotation periods of around 171 days (the value we have adopted in this work) based on Gaussian Process regression analyses of spectropolarimetric data while \citet{Shan2024} report a rotation period of 53 days based on a generalised lomb scargle periodogram analysis of $R'_{\rm HK}$ and TiO time series. If Gl 408's rotation period is actually 53 days, this would reduce its Rossby number by a factor of $\sim$3, possibly shifting the position of this star in fig. \ref{fig:BVsRo} from the population of slowly rotating FC stars (square points) to the population of unsaturated PC stars (circular points around the red best fit line). This would then resolve the oddity of single a PC star (Gl 408) having the same magnetic properties as the slowly rotating FC stars. However, we have adopted the 171 day value over the 53 day value for consistency since it is the rotation period used by \citet{Lehmann2024} to produce the ZDI map for Gl 408 even though this is an unusually long rotation period when compared to other stars of similar masses to Gl 408 \citep[e.g.][]{Newton2018}.

For now, the true period of Gl 408 remains unclear as well as whether the change in magnetic field properties occurs exactly at the FC boundary or if a parameter other than stellar structure is playing a role. Further ZDI observations of stars around the FC boundary are needed to clarify exactly what is causing the change in magnetic field properties in this area of parameter space.

\subsection{Convective turnover times}
\label{subsec:TurnoverTimesDiscussion}
The convective turnover time is a notoriously difficult parameter to estimate with many prescriptions existing in the literature. One might, therefore, ask if the fact that the slowly rotating FC stars in our sample do not follow the same Rossby number scaling as the PC stars is because of our choice of turnover time prescription. Perhaps an alternative prescription would erase this result? Looking at, e.g. fig. \ref{fig:BVsRo}, the Rossby number values of the slowly rotating FC M dwarfs would need to be decreased by a factor of around 5 to 10 to bring them in line with the PC stars. In other words, their turnover times would need to be increased by about a factor of 5 to 10 while the turnover times of the PC stars are left unchanged. 

In fig. \ref{fig:Tau}, we have plotted some recent published turnover time prescriptions that display a sharp rise in turnover times at the FC boundary \citep{Chiti2024,Gossage2024}. These prescriptions all look broadly similar to our own prescription, equation (\ref{eq:TauPres}). While some of them predict a slightly bigger rise in turnover times at the FC boundary than our prescription, none predict turnover times for FC stars that would be large enough to bring the FC stars in fig. \ref{fig:BVsRo} in line with the PC stars. Indeed, to our knowledge, there are no prescriptions in the literature that would bring the FC stars in our sample in line with the PC stars. Therefore, our results appear to be real, rather than a consequence of our choice of turnover time prescription.

\subsection{Rotation evolution}
\label{subsec:RotEvo}
Our findings have significant implications for rotation evolution modelling. Typically, in rotation evolution models, the same magnetic field strength vs Rossby number scaling is used for both FC and PC stars \citep[e.g.][]{vanSaders2013,Matt2015,Johnstone2021}. However, it is now clear that the magnetic fields of FC and PC stars have different Rossby number scalings in the unsaturated regime and future rotation evolution models will need to account for this. What remains uncertain is the magnetic properties of FC M dwarfs with intermediate rotation periods. For example, do FC stars with intermediate rotation periods also have stronger large-scale magnetic fields than their PC counterparts or not? While ZDI has been used to probe the magnetic properties of rapidly rotating and slowly rotating FC M dwarfs (see fig. \ref{fig:Params}), it has yet to be used to reconstruct the magnetic fields of intermediate period FC M dwarfs. Until this is done, rotation evolution models for FC M dwarfs will be missing an important constraint.

\section{Conclusions}
\label{sec:Conclusion}
In this work, we have assembled a sample of 96 stars that have been collectively mapped 260 times with ZDI. With this sample, we have analysed the magnetic and spin-down properties of main sequence stars over a wide range of masses and rotation periods. We have paid special attention to the slowly rotating fully convective stars in our sample as ZDI maps for these types of stars have not been available until relatively recently \citep{Klein2021,Lehmann2024}.

Similar to previous studies, we find that the magnetic fields of the majority of our sample are well parameterised by the Rossby number with clear saturated and unsaturated regimes being evident. However, the slowly rotating FC stars have significantly stronger magnetic fields than expected given their Rossby number (section \ref{subsec:ActRotRel}). This behavior is seen in the total large-scale magnetic field strength as well as the field strengths of individual components, i.e. the dipole, quadrupole or octupole. This suggests that different mechanisms are operating in the dynamos of FC and PC stars resulting in the different large-scale magnetic field strengths of FC and PC stars, for a given Rossby number (section \ref{subsec:Dynamo}). However, the dynamo mechanisms generating small-scale fields appear to be similar in FC and PC stars since X-ray emission, a probe of small-scale magnetic fields, scales in the same way as a function of Rossby number for both types of stars \citep{Gossage2024}. One consequence of the differing large-scale magnetic field strengths is that slowly rotating FC stars are expected to lose angular momentum, via their magnetised winds, significantly faster than PC stars with comparable Rossby numbers (section \ref{sec:Spindown}).

We have also investigated how the toroidal and poloidal field components varies across our sample of stars. In toroidal energy density vs poloidal energy density parameter space, our stars lie on two, almost parallel sequences with stars more massive than $\sim 0.5 M_\odot$ forming one sequence and stars less massive than $\sim 0.5 M_\odot$ forming the other. While we had already previously studied the poloidal and toroidal magnetic fields in \citet{See2015}, the presence of the two sequences is much clearer in this work due to the increased sample size. Their existence, however, is difficult to explain. It is unclear what the significance of the stellar mass 0.5$M_\odot$ is and why it should be the division point of the two sequences.

\section*{Acknowledgements}
We thank the referee for comments that helped improve this manuscript. VS acknowledges support from the European Space Agency (ESA) as an ESA Research Fellow and from the European Research Council (ERC) under the European Union’s Horizon 2020 research and innovation programme (CartographY G.A. n. 804752). AAV acknowledges funding from the European Research Council (ERC) under the European Union's Horizon 2020 research and innovation programme (grant agreement No 817540, ASTROFLOW). AAV and SB acknowledge funding from the project ``Exo-space weather and contemporaneous signatures of star-planet interactions" (with project number OCENW.M.22.215 of the research programme ``Open Competition Domain Science - M"), which is financed by the Dutch Research Council (NWO). BK acknowledges funding from the European Research Council under the European Union’s Horizon 2020 research and innovation programme (grant agreement no 865624, GPRV). JFD and LTL acknowledged funding from the European Research Council (ERC) under the H2020 research \& innovation programme (grant agreement \#740651 NewWorlds). CPF acknowledges funding from the European Union's Horizon Europe research and innovation programme under grant agreement No. 101079231 (EXOHOST), and from the United Kingdom Research and Innovation (UKRI) Horizon Europe Guarantee Scheme (grant number 10051045). LA acknowledges support from the Centre National des Etudes Spatialees (CNES) through the PLATO/AIM grant. AJF acknowledges funding from the European Research Council (ERC) under the European Union’s Horizon 2020 research and innovation programme (grant agreement No 810218 WHOLESUN). MMJ acknowledges support from STFC consolidated grant number ST/R000824/1.

\emph{Software:} \texttt{matplotlib} \citep{Hunter2007}, \texttt{numpy} \citep{Harris2020}, \texttt{scipy} \citep{Virtanen2020}

\section*{Data Availability}

The numerical data used in this study, i.e. those given in table \ref{tab:Params} will be available online in machine readable format upon publication.



\bibliographystyle{mnras}
\bibliography{Main} 

\begin{figure*}
    \includegraphics[trim=0mm 15mm 0mm 0mm,width=\textwidth]{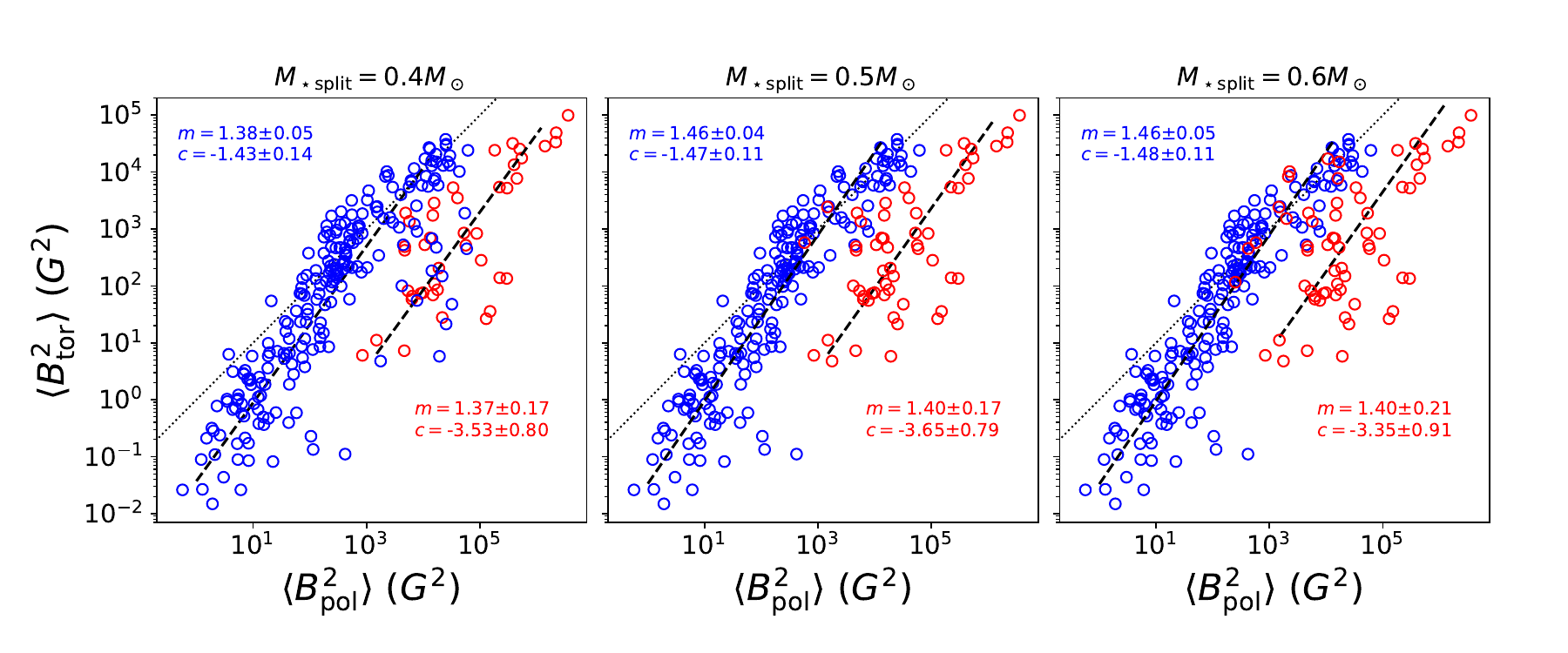}
    \caption{Toroidal magnetic energy density vs poloidal magnetic energy density for our sample stars. In each panel, stars are divided into two sub-samples based on their stellar mass. The stellar mass value used to define the two sub-samples, denoted $M_{\rm \star split}$, is varied and shown above each panel. Stars with $M_\star \geq M_{\rm \star split}$ are shown in blue and stars with $M_\star < M_{\rm \star split}$ are shown in red. For each choice of $M_{\star \rm split}$, we perform fits of the form $\log_{10} \langle B_{\rm tor}^2\rangle = {\rm m} \log_{10} \langle B_{\rm pol}^2\rangle + {\rm c}$ for both sequences. The best fit values of $\rm m$ and $\rm c$ are shown in each panel and the corresponding fits are shown as dashed black lines.}
    \label{fig:MassSplit}
\end{figure*}

\begin{figure*}
    \includegraphics[trim=0mm 5mm 0mm 0mm,width=0.6\textwidth]{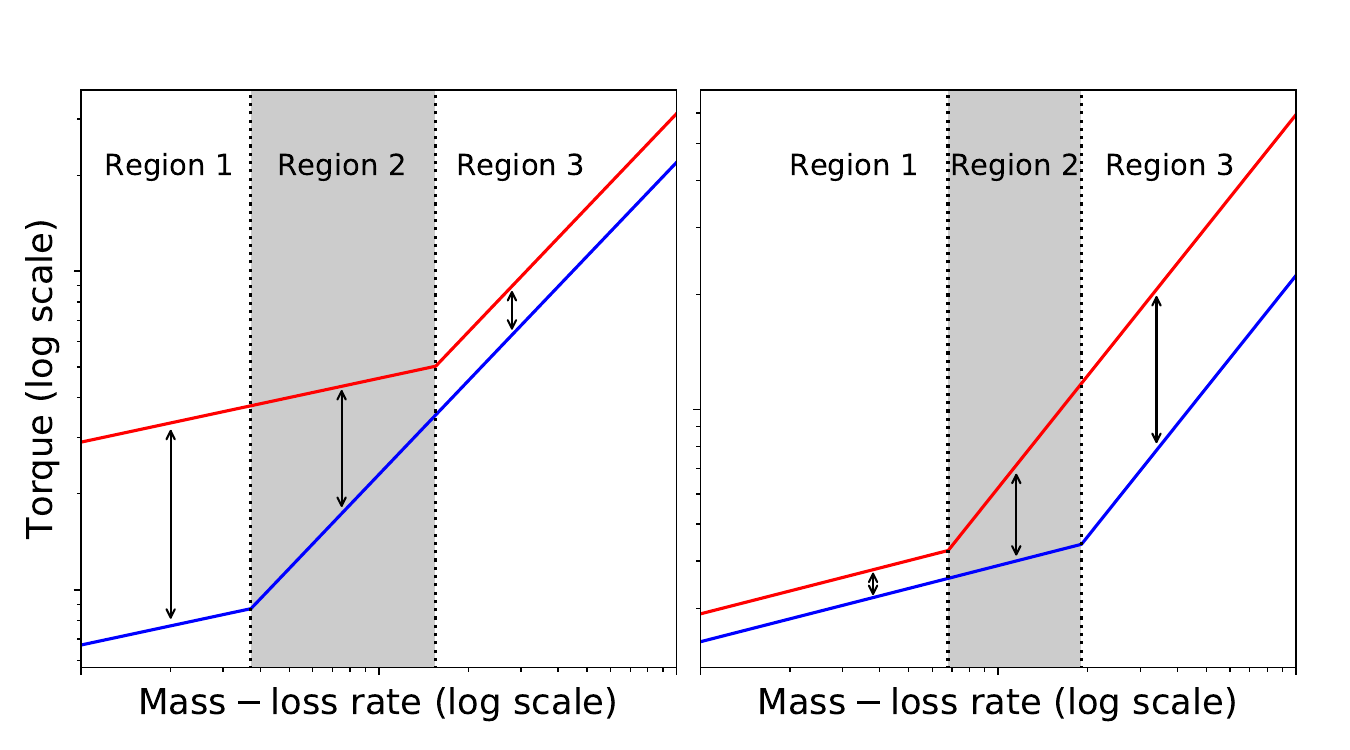}
    \caption{A sketch of how the torque varies as a function of mass-loss rate (see equation (\ref{eq:Torque}) and (\ref{eq:rA})) when only considering dipole and quadrupole field components. The red and blue curves represent $T_{\rm pred}$ and $T_\star$ respectively. The torque ratio, as defined in section \ref{sec:Spindown}, is proportional to the vertical distance between the red and blue curves, with examples shown by the arrows. Three regions can be identified that are defined by where the breaks in the red and blue curves occur (dotted lines). The left and right panels show two different examples of $T_{\rm pred}$ and $T_\star$.}
    \label{fig:TorqueRatioSketch}
\end{figure*}


\appendix
\section{What mass best separates the two sequences in poloidal-toroidal magnetic energy space?}
\label{sec:MassSplit}

In section \ref{subsec:PolTor}, we showed the toroidal magnetic energy density vs the poloidal magnetic energy density of our sample (fig. \ref{fig:TorVsPol}). We noted that two sequences are evident in this plot, one composed of higher mass stars and the other of lower mass stars. However, the stellar mass value that best separates the two sequences is not clear cut. For the purposes of this appendix, we denote this mass as $M_{\star \rm split}$. Although we have used $M_{\star \rm split}=0.5M_\odot$ throughout this work, it is worth exploring different choices of $M_{\star \rm split}$, which we do in this section.

In fig. \ref{fig:MassSplit}, we show the toroidal magnetic energy density vs the poloidal magnetic energy density of our sample again. Stars with $M_\star \geq M_{\rm \star split}$ are shown in blue and stars with $M_\star < M_{\rm \star split}$ are shown in red. Each panel corresponds to a different value of $M_{\rm \star split}$, either 0.4$M_\odot$, 0.5$M_\odot$ or 0.6$M_\odot$. For $M_{\star \rm split}=0.4M_\odot$ (left panel), we see that there is a considerable scatter of points from the higher mass sequence (blue points) into the bulk of the lower mass sequence (red points). Conversely, for $M_{\star \rm split}=0.6M_\odot$ (right panel), points from the lower mass sequence scatter into the main trend of the higher mass sequence. For the intermediate value of $M_{\star \rm split}=0.5M_\odot$, there is still some scatter from the lower mass sequence into the higher mass sequence but considerably less than for the $M_{\star \rm split}=0.6M_\odot$ case. As such, $0.5M_\odot$ seems to be the stellar mass value that best delineates the two sequences. It is worth noting that there are only 21 ZDI maps in our sample belonging to stars with masses between 0.4$M_\odot$ and 0.6$M_\odot$ which is relatively few stars. While it seems fairly evident that $M_{\star \rm split}$ is somewhere in this range of masses, further observations of stars in this mass range would allow us to better determine the value of $M_{\star \rm split}$. We also note that, regardless of our choice of $M_{\star \rm split}$ within this mass range, the best fit lines for the two sequences are the same within the error bars.

\section{``Deriving'' the torque ratio expression}
\label{sec:TorqueRatioExplain}
In section \ref{sec:Spindown}, we stated that the true value of the torque ratio lies somewhere within a range of values as given by equation (\ref{eq:TorqueRatioRange}). In this section, we explain where this equation comes from. The mathematical derivation of equation (\ref{eq:TorqueRatioRange}) is rather long and not particularly instructive in, and of, itself. As such, it is simpler, and more useful to the reader, to ``derive'' equation (\ref{eq:TorqueRatioRange}) visually.

The reason why we can only determine a range of values rather than the true value of the torque ratio is because the mass-loss rates of our stars are unknown. As a reminder, the braking law of \citet{Finley2018} shown in equations (\ref{eq:Torque}) and (\ref{eq:rA}) is a twice broken power law, as a function of mass-loss rate, with the different segments of the broken power law being associated with the dipole, the dipole + quadrupole, and the dipole + quadrupole + octupole components respectively. However, to begin with, it will be simpler to ignore the octupole component such that the braking law is given by a once broken power law. 

To reiterate, our goal is to determine the value of the torque ratio, $T_{\rm pred}/T_\star$, i.e. what is ratio of toques for two stars, one real and one hypothetical, that have the same parameters except for their magnetic field strengths. In the left hand panel of fig. \ref{fig:TorqueRatioSketch}, we show a sketch of how the torque varies as a function of mass-loss rate for two stars in red and blue respectively using the braking law of \citet{Finley2018}, ignoring the octupole field component. We can, arbitrarily, assign the red curve to be $T_{\rm pred}$ and the blue curve to be $T_\star$. For the purposes of this explanation, the values of the stellar mass, radius, rotation and mass-loss rate are not important as long as they are the same for $T_{\rm pred}$ and $T_\star$. The values of the torque and mass-loss rates along the x and y axes are also unimportant although it is worth highlighting that these plots are on a logarithmic scale for both axes. The difference between the red and blue curves is solely because they have different dipole and quadrupole field strengths. Each curve is composed of a shallower segment at smaller mass-loss rates and a steeper segment at larger mass-loss rates. The shallower segment is associated with just the dipole component of the field and is given by equations (\ref{eq:Torque}) and (\ref{eq:rADip}) while the steeper segment is associated with the dipole + quadrupole components and is given by equations (\ref{eq:Torque}) and (\ref{eq:rAQuad}). We note that the shallower segments of both curves have the same slope and similarly, the steeper segments of both curves have the same slope.

Figure \ref{fig:TorqueRatioSketch} can be split into three regions (which are labeled in the plot), defined by where the breaks in the blue and red curves occur (these are denoted by the vertical dotted lines). The value of the torque ratio depends on the region that a star is in, which is determined by the (unknown) mass-loss rate. Since fig. \ref{fig:TorqueRatioSketch} is on a logarithmic scale, the torque ratio, $T_{\rm pred}/T_\star$, is proportional to the vertical distance between the two curves. In region 1, both the red and blue curves are in the dipole dominated domain. As such, the torque ratio has a fixed value given by equation (\ref{eq:TorqueRatioDip}) for any mass-loss rate in this region. Similarly, in region 3, both the red and blue curves are in the dipole + quadrupole domain and the torque ratio has a fixed value given by equation (\ref{eq:TorqueRatioQuad}) for any mass-loss rate in this region. In region 2, the red curve is in the dipole regime while the blue curve is in the dipole + quadrupole regime. As such, the torque ratio does not have a fixed value in this region and depends on the mass-loss rate. From a visual inspection, one can see that, regardless of the mass-loss rate, the torque ratio in region 2 is smaller than the torque ratio in region 1 but bigger than the torque ratio in region 3.

For the example shown in the left panel of fig. \ref{fig:TorqueRatioSketch}, our choice of dipole and quadrupole field strengths for the red and blue curves has resulted in the torque ratio in region 1 being bigger than the torque ratio in region 3. However, a different choice of field strengths could have resulted in the reverse being true which is shown in the right panel of fig. \ref{fig:TorqueRatioSketch}. In this case, the torque ratio in region 2 is bigger than the torque ratio in region 1 but smaller than the torque ratio in region 3. Therefore, the general rule (when neglecting the octupole component) is that the extrema values that the torque ratio can take are given by the torque ratios in the dipole only domain (region 1, equation (\ref{eq:TorqueRatioDip})) and the dipole + quadrupole domain (region 3, equation (\ref{eq:TorqueRatioQuad})) with all values in between being possible depending on the mass-loss rate. Mathematically, this can be stated as 

\begin{equation}
\begin{split}
    {\rm min}\left( \frac{T_{\rm pred}}{T_\star} \right) &= {\rm min} \left\{ \frac{T_{\rm pred}}{T_\star}\bigg\rvert_{\rm d}, \frac{T_{\rm pred}}{T_\star}\bigg\rvert_{\rm q} \right\}, \\
    {\rm max}\left( \frac{T_{\rm pred}}{T_\star} \right) &= {\rm max} \left\{ \frac{T_{\rm pred}}{T_\star}\bigg\rvert_{\rm d}, \frac{T_{\rm pred}}{T_\star}\bigg\rvert_{\rm q} \right\},
    \label{eq:TorqueRatioRangeAppendix} 
\end{split}
\end{equation}
where the symbols in this equation have the same meanings as in section \ref{sec:Spindown}.

Although we have only considered the dipole and quadrupole components so far, the same arguments can be extended to include the octupole field component. Equation (\ref{eq:TorqueRatioRangeAppendix}) would then be extended to include an additional term, $\left. \frac{T_{\rm pred}}{T_\star}\right\rvert_{\rm o}$ in the minimum and maximum functions which is just the expression given in equation (\ref{eq:TorqueRatioRange}).


\bsp	
\label{lastpage}
\end{document}